# Fault Detection for Non-Condensing Boilers using
# Simulated Building Automation System Data

Rony Shohet, Mohammed Kandil, J.J. McArthur*

Faculty of Engineering and Architectural Science, Ryerson University, Toronto, Canada

* Corresponding Author: 325 Church Street, Toronto, ON M5B 2K3, Canada; Tel: 1 416 979 5000 x554082; Fax: 1 416 979 5353; jjmcarthur@ryerson.ca.

## Abstract

Building performance has been shown to degrade significantly after commissioning, resulting in increased energy consumption and associated greenhouse gas emissions. Fault Detection and Diagnosis (FDD) protocols using existing sensor networks and IoT devices has the potential to minimize this waste by continually identifying system degradation and re-tuning control strategies to adapt to real building performance. Due to its significant contribution to greenhouse gas emissions, the performance of gas boiler systems for building heating is critical. A review of boiler performance studies has been used to develop a set of common faults and degraded performance conditions, which have been integrated into a MATLAB/Simscape emulator. This resulted in a labeled dataset with approximately 10,000 simulations of steady-state performance for each of 14 non-condensing boilers. The collected data is used for training and testing fault classification using K-nearest neighbour, Decision tree, Random Forest, and Support Vector Machines. The results show that the Decision Tree, Random Forest, and Support Vector Machines method provide high prediction accuracy, consistently exceeding 95%, and generalization across multiple boilers is not possible due to low classification accuracy.

## 1    Introduction

A building's performance may range from several technical perspectives, including structural performance, mechanical performance, envelope performance, and architectural performance [1].

The discussion of building performance within the context of this research is primarily focused on the performance of heating, ventilation, and air condition (HVAC) systems. HVAC play a fundamental role in maintaining comfortable indoor environments. However, their operation is not always perfect, where HVAC equipment faults can inhibit performance through various means. These faults include the degradation of equipment, failing of sensors, improper equipment installation, poor maintenance, and improperly implemented controls [2]. These HVAC systems can be complex, and while streamed data is available, physical inspection – particularly of internal components – is often not possible, making fault detection and diagnosis a difficult task for building operators. The accumulation of faults can lead to increased energy costs, a loss of comfort causing a loss of productivity or loss of tenants, or increased wear of components, or decreasing reliability of equipment [2]. The impacts of these faults include approximately 15 to 30% waste of energy [3].

Building Automation Systems (BAS) provide centralized control for most building operations as well as sensor point monitoring, which can support fault detection and diagnosis (FDD). Collected sensor data provides an opportunity to develop a tool capable of evaluating performance of the observed physical system through FDD [4]. FDD is an effective tool for evaluating system performance and has been broadly used in the automotive, nuclear, and aerospace sectors [4]. This approach has been more recently adopted within the building sector [4] because efficiency losses due to mechanical equipment degradation and poor controls can be significant and are well-addressed by this approach [5]. Therefore, FDD allows for improved maintenance opportunities through increased insight on building performance.

A physics-based model of a non-condensing boiler is developed in MATLAB/Simscape. The primary purpose of this model is to supplement the lack of actual sensor data. In addition, this model is capable of replicating faults that may occur within the boiler. Generating fault data using simulation is extremely valuable as it permits the modeling for complex fault scenarios (multiple concurrent faults) and is a way to inexpensively generate the bulk data necessary for algorithm development and testing [6]. Further, this approach permits data on rare or dangerous fault conditions to be generated without risk to the building or its occupants.

This paper focuses on the development of an FDD tool to permit buildings with Building Automation Systems to diagnose heating system faults; such buildings typically use hydronic heating systems consisting of boilers and circulation pumps. Because detailed fault data is not available from in-situ HVAC equipment, a simulated dataset was required. To develop this, a series of boiler emulators were created based on thermodynamic relationships modelled in MATLAB/Simscape which are then calibrated and validated according to manufacturer test data in nominal conditions. These were then used to create large labelled datasets of simulated nominal and fault condition data. Finally, different machine learning algorithms were used to develop a process-history based FDD classifier using only data available from typical BAS systems.

This paper contributes to the development of improved FDD in several ways. First, it presents a set of physics-based boiler models calibrated to manufacturer data that can be used to generate simulated data for rare fault conditions. The research acted as a case study that evaluated the performance of a developed FDD methodology across 14 boiler models. Second, this paper evaluates a range of potential machine learning algorithms for FDD development and identifies the algorithms providing very high accuracy (average of 95%). Finally, a generalization study has been conducted to determine whether a single FDD classifier can be used across a range of non-condensing boiler models. Collectively these contributions create the theoretical framework necessary to perform accurate FDD for hydronic heating systems using data generated from physics-based models.

## 2   Literature Review

The modelling of physical systems and the application of FDD techniques has been researched extensively within the HVAC industry. This FDD tool was developed by first creating a physics-based (physical, or "white box") model of the investigated equipment. This modelling provides the foundation for understanding equipment behaviour and how its performance changes due to the presence of faults. The published literature of the physics-based models of HVAC equipment, specifically those applied to heating system are discussed in the following sections. Other modelling methods includes both data-driven (black box) and gray box (hybrid) models. Both are briefly reviewed, summarizing their theory and published literature. In addition, the underlying physics of faults are investigated since most physics-based models are assumed to operate under normal conditions. The goal of which is to emulate the faults that occur within boilers, all the while generating data that replicates performance degradation. The literature review then discusses the application of physics-based models towards FDD development. FDD is typically applied to reduce equipment downtime, while making maintenance, repair, and prevention efforts as efficient as possible [1]. This is achieved by actively identifying outliers within building behaviour that indicate whether a fault exists [1].

### 2.1 Physics-based Modelling Methods for HVAC Equipment

Physics-based models are digital twins of HVAC equipment, including inputs and outputs as well as the thermodynamic, fluid-dynamics, and heat-transfer behaviour of equipment and its component parts to allow emulation of equipment performance. The strength of white-box models lies in their ability to be generalized [7, 8, 9]. Unlike data-driven ("black-box") models that are limited to the data domain used for training, white-box models can encompass a wider range of operating conditions, providing users with an increased understanding of how performance changes. Finally, the white-box model represents a system more closely as it uses specific knowledge of the inner workings of a system [7].

Physics-based models using a white-box modelling strategy have been applied for a wide range of HVAC equipment, including Air Handling Units (AHU), cooling and heating coils, fan and pumps, chillers, and cooling towers [7, 8]. Many physics-based models of boilers have been developed and vary according to their level of detail, the calculation method, and input parameters [10]. Glembin et. al. [10] provides an overview of developed boilers models. The authors present a series of published literature that models boilers as steady-state systems. Many of the discussed boiler models conceptualize the boiler as several components that interact with each other. These components include the combustion chamber and the combustion gas – water heat exchanger. However, the complexity of modelling those components vary. Firstly, the combustion chamber is described as mostly modelled under adiabatic conditions. This means that heat and/or mass is not exchanged with its surroundings, staying within the system boundaries [11]. Meanwhile, the combustion gas – water heat exchanger is typically modelled as a counterflow heat exchanger. The outlet temperatures of the heat exchanger are found using an effectiveness-NTU method. Several of the discussed papers include additional heat exchangers that model the exhaust (flue) gas temperature, heat loss to surroundings, or water condensation of condensing exhaust gas. In addition to the discussed steady-state models, the author presents a summary of dynamic models. These models are capable of simulating performance with varying input conditions throughout the simulated timestep. The discussed models capture a higher level of detail with regards to actual boiler performance. Boilers are not typically operated at steady-state conditions, rather the boiler cycles or modulates or input values change, highlighting the importance of dynamic modelling [10]. In addition, Glembin et. al. discussed the development of a low complexity dynamic boiler model. Complexity is minimized by utilizing a low number of input parameters, taken mostly from manufacture data. The model is programmed using Fortran within the TRNSYS environment. Steady-state performance is modelled using the two-component method, where the interactions between a combustion chamber and gas-water heat exchanger are presented. While the dynamic model performs calculation of flue gasses and water side temperatures using differential equations and time constants. The dynamic boiler model proves to model steady-state performance highly accurately, while dynamic performance suffered from poor accuracy.

Additional physics-based models of boilers include Satyavada and Baldi [12], who developed a dynamic condensing boiler model within MATLAB/Simulink. The boiler model formulates a set of 4 partial differential equations that model condensing and non-condensing boiler operation upon infinitesimal elements. The four equations solve the time varying temperature values of the water, gas, surface temperature, and the tube wall temperature. However, if condensation occurs within the heat exchanger the temperature of the flue gas remains constant, meaning the set of partial differential equations becomes 3. The results of the boiler model conform to published manufacturer data and several other published boiler models. Furthermore, Haller et al. [13] investigated the steady-state modelling of multiple fuels, including oil, gas, pellets and wood chips. The authors present 3 methods including an empirical delta-T approach, empirical effectiveness approach, and detailed effectiveness-NTU method, all developed with Fortran for use in TRNSYS.

The results were most accurate with the effectiveness-NTU method; however, the authors note the limited information published by manufacturers may result in poor accuracy. Ternoveanu and Ngendakumana [14] present a dynamic boiler model that models start-up, cool-down, and steady-state performance. The methodology of steady-state performance is consistent with other papers. Two primary components are modelled: a combustion chamber and gas-water heat exchanger. Flue gas and water temperatures are found by relating inputs and outputs with energy balance equations. The results of the paper were promising for modelling steady-state performance. While, the model suffered from an inability to accurately calculate flue gas temperatures during start-up and cool-down. Makaire and Ngendakumana [15] present a methodology for the mathematical modelling of a condensing boiler operating at steady-state conditions. The model is broken into three components, a combustion chamber, dry gas-water heat exchanger, and wet gas-water heat exchange. Combustion is modelled assuming adiabatic conditions and complete combustion. The dry heat exchanger is modelled with an effectiveness NTU methodology, assuming counterflow between combustion gasses and water. Finally, the wet heat exchanger, which models condensation of flue gasses, is assumed to occur as if wet air were passing over a cooling coil. This model was applied towards two residential boilers, one using gas and the second using oil. It is required to calculate 6 parameters that fit the boiler model to the measured boiler. The 6 parameters generally encompass nominal flow rate, and heat transfer coefficients for the two heat exchangers. It was found that the models were able to accurately calculate boiler efficiency across multiple inlet water temperatures. However, this model used data from an actual boiler therefore making it a gray box model. Finally, Aganovic [16] presents the methodology of modelling a boiler room, including the boiler, pumps, pipes, and valves using the Modelica language, Dymola, and the Buildings Library for HVAC components by Simulation and Research Group at Lawrence Berkley National Laboratory. Using the Buildings Library, the author develops a model of dynamic boiler operation. Combustion is not modelled directly, rather efficiency curves are used that relate return water temperature and nominal power to calculate the heat of combustion of the fuel. This is then input to the boiler control volume, that calculate the heat gain towards supply water. The results are evaluated on an hourly basis, proving good agreement between measured and simulated behaviour.

The white-box modelling of HVAC systems proves to be complex due to the interactions across multiple domains, including, thermodynamics, fluid mechanics, heat transfer, electro-magnetic, control systems, and electrical circuits [17]. Fortunately, there exist software capable of white-box modelling of HVAC equipment, including Modelica [18], Simscape [19], TRNSYS [20], HVACSim+ [21]. These modelling software handle many of the cumbersome differential equations present in HVAC components. This paper utilizes Simscape, a toolbox within MATLAB/Simulink for several reason. First, MATLAB, with or without Simulink, is broadly used for model predictive control, particularly the development of physics-based models to simulate system behavior. In addition, Simscape allows for the modelling of physical systems across multiple domains including electrical, mechanical, thermal, thermal liquid, two-phase liquid, gas, and moist air. Simscape avoids abstract mathematical equations through the application of component blocks, providing the designer with a user-friendly and simple design process. Each block represents a component of a physical system, allowing for multiple components to be combined through their respective energy flows. As for the design of the non-condensing boiler, the structure was inspired by the work presented by Mathworks [22]. This example demonstrates the application of a boiler for the heating of a residential house. The components of the system – combustion chamber, heat exchanger, and a water source – were adopted for this research. The fundamental component of the combustion chamber was alleviated by this example [22]. The provided component could emulate the heat and mass transfer associated with combustion for any hydrocarbon fuels. The parameters within the heat exchanger component were modified for each boiler model according to manufacturer data. In addition, this paper leveraged the provided components to provide data on fault scenarios, including, increasing excess air within the combustion chamber, and increasing fouling and scaling within the heat exchanger. The methodology of the parameter selection and fault modelling is described further in this paper.

As for examples that use Simscape within literature include, Lapusan et. al. [23] and Behravan et. al. [24] who both model the thermal dynamics of a heating system within a building, while avoiding direct simulation of a boiler. The heating source is designated as either a lumped model [23], or a heat source adjusted according to the necessary heat output required to achieve a temperature setpoint [24]. Pavlusova et. al. [25] utilizes Simscape to model the performance of a boiler within a multizone residential building. The authors model the boiler manually by applying differential equations that calculate the heat generated. However, this task can be simplified by utilizing the heat exchanger component provided within Simscape. Dahash et. al. [26] present a comparison between a EBILSON, a Modelica based simulation software, and a Simscape model for district heating system model. The research models a heat load, heat source, piping network and pumps with first principles. The heat source is simplified to model the dynamics of the water loop and assumes constant heat inputs. This research concludes that despite Simscape having less literature discussing its application towards physics-based modelling, it is as capable as Modelica in modelling complex physics-based systems [26].

## 2.2 Other Equipment Modelling Approaches

Two other modelling techniques exist, namely black-box and gray-box models. Black-box models rely on collected field data to characterize equipment performance and have the advantage of requiring little knowledge of the system and its processes, allowing for accurate representation of the system [8]. Although such a model is only as accurate as the training data allows it to be -- if new data is presented that is sufficiently different from the training data, the model suffers from poor accuracy [8]. This disadvantage is highlighted when performance of equipment deviates due to presence of faults, thus they must be regularly updated to enable modelling of operation and performance changes [27].

Jamshidi et. al. present a black-box model of steam boiler using Takagi-Sugeno fuzzy model [28]. Using only a single input (fuel gas flow rate) and single output (steam temperature) collected from experimental data, the authors were able to replicate boiler performance with a high degree of accuracy. Desai et. al. presents a multiple input and multiple output black box for a drum type steam boiler [29]. The authors developed several transfer functions capable of modelling components within the steam boiler, including pump, control valves, pressure, and water levels. Finally, a high efficiency boiler is modelled using neural networks within a multiple input and single output black-box model. The model is designed to determine the outlet temperature of the water based on the inputs of inlet water temperature, water flow rate, and the load of the boiler [30]. The results on untrained data showed significant errors, likely due to poor model training. The general development of the model was said to be fast, simple, and did not require any a priori knowledge of the system.

Gray-box modelling uses a combination of both black and white-box techniques [8, 27]. Several examples of published literature of gray-box models of HVAC equipment is presented. Firstly, Farooq et al. [31] presented a methodology for modelling a low-pressure electric hot water boiler. The authors first modelled the stratified water temperature with 8 energy balance equations in Simulink. While the parameters were evaluated from collected data using non-linear least squares optimization. The results of two cases of boiler performance were replicated with high levels of accuracy. Rusinowski and Stanek created a hybrid model of an industrial steam boiler to evaluate the reduction in efficiency due to flue gas losses and unburnt combustible losses [32]. Energy balance equations are formulated such that the energy fluxes across the boiler are accounted for. A neural net and regression model are used to evaluate flue gas temperatures from 6 inputs. This model can calculate the efficiency due to losses from measurements of flue gas composition.

## 2.3   FDD Methods for HVAC Equipment

FDD requires an understanding of faulty equipment behaviour and a means to classify whether observed behaviour is, in fact, showing equipment to be in a fault condition. A fault has been defined as any deviation from the acceptable range of an observed variable [33]. Generally, FDD is described as a three-step process: fault detection, fault isolation, and fault identification [4]. Fault detection determines whether any abnormal operation is occurring through measurements of sensor data. Fault isolation determines the location of the fault and the time of detection. Fault identification determines the nature of the fault and its significance in terms of impact on the overall system. These impacts can be quantified in terms of safety, costs, energy, comfort, or emissions. As well, an automated FDD scheme includes a *decision*, that may signal an alarm, shut down the system, reconfigure the controls, or notify facility management of necessary repairs. The primary goal of FDD is the early detection and diagnosis of faults, allowing for correction that would prevent further damage if left untreated. Therefore, often FDD applications require the observed system to be continuously monitored. A generic process of FDD has been discussed; however, there are many methods for the successful application of FDD.

The various methods of FDD are broadly categorized as quantitative model-based, qualitative model-based, and process history-based [4, 33, 34, 35]. Figure 1 depicts a detailed depiction of the types of FDD methods that have been used within maintenance of HVAC equipment. It is noted that despite the generic classification of FDD methods, these are not strictly for HVAC equipment. Examples of FDD applications within other aspects of building performance include whole building systems, as described in [36].

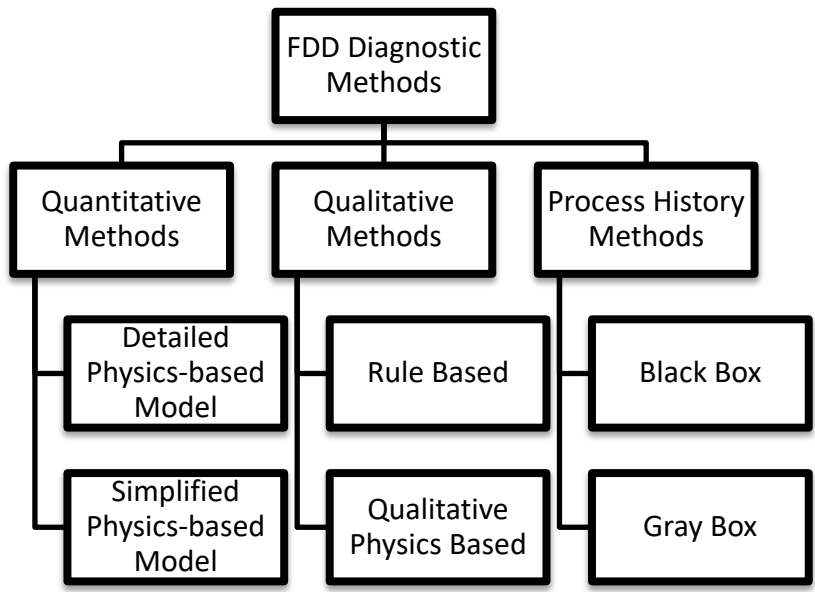

*Figure 1 Classification Scheme of FDD Methods [4]*

Both quantitative and qualitative model-based FDD utilize an understanding of the underlying physics of the system to develop relationships between inputs and outputs [4, 33]. However, quantitative model-based FDD creates a relationships between inputs and outputs from mathematical functions [4, 33]. While, a qualitative model-based FDD expresses the relationship through a qualitative understanding of the physics-based system [4, 33]. Meanwhile, the biggest difference between model-based approaches and the process history approach is the source of a priori knowledge used to evaluate faults. In model-based or white-box approaches [33], this knowledge is gathered from first principle relationships, whereas process history-based methods (black box approaches) gather this information from historical data correlations with no regard to the system physics [33]. Hybrid approaches combine both models and historical data using a gray-box approach. The following sections further discusses the background of FDD, and the various applications of FDD for HVAC maintenance.

### 2.3.1 Boiler FDD

It is assumed model-based methods have a lack of sensor data; therefore, the parameters are often developed through an understanding of the underlying physics of the process. The quantitative model-based approach uses mathematical functions to define relationships between inputs and outputs of systems. Analytical redundancy is used to compare sensor measurements to model outputs ('analytically measured results') and uses the differences ('residuals') as inputs to the FDD algorithm [4]. Meanwhile, the qualitative model-based approach also uses a physical understanding of the system while observing qualitative relationships between inputs and outputs. For example, an expert system, where a computer mimics the problem-solving methodology that a human expert would follow in order to come to a conclusion [4]. The goal of qualitative model-based is to encompass a detailed understanding of the system while avoiding the rigidity of numeric solutions that often overfit systems [34]. The developed physics-based model provides simulated sensor data across a range of operating conditions, including faults. This indicates that the FDD methodology is a process-history method.

The challenge with using a physics-based models for FDD includes the simulation of faults. As noted by Li and O'Neill [6], most physics-based models of HVAC equipment assume normal operation with no failure points. The modelling of faults is a non-trivial task, requiring expertise in understanding their impact, all the while requiring a fundamental understanding of mathematically modelling the faults. Some papers discuss a fault detection methodology based on modelled boiler performance [37]. Typically, the methodology uses a residual computation, meaning the model calculates normal (fault-free) behaviour and compares the calculated value to a measured sensor value. If a residual exists, it means there is deviation from normal performance that the model calculates and therefore the actual measured equipment would contain a fault. Additional papers that implement fault modelling include Haves [38], who presents a methodology for simulating faults within a cooling coil. The authors describe the development of fault models through two means. The first being changes in parameters of modelled equipment, for example, reducing UA values to simulate fouling in a heat exchanger. The second method includes a detailed approach that explicitly models faults as new parameters that define the severity of the faults. A common conclusion of implementing fault modelling is the challenge of validating fault models without collected data, specifically data of faulty performance. But collecting this validation data is a time-consuming task because it takes time for equipment to fail. The time-consuming nature of data collection justifies the need for developing fault models. An added benefit of fault modelling includes verifying the robustness of an FDD scheme by evaluating many models of faulty HVAC equipment.

Examples of boiler FDD includes the work by Baldi et al. [39] who performed model-based FDD on a condensing boiler model. The authors develop a strategy capable of detecting deviations of boiler efficiency and faults for sensors and actuator. This work also presents the novel idea of "virtual sensors", a method for estimating flow rates, that avoids the cost of installing expensive flow meters. The application of model-based FDD schemes has seen various non-heating HVAC applications. Liang and Du [3] developed a model-based FDD of an entire HVAC system – chiller, air handling unit, cooling coils, ducting, dampers, air mixers, and the thermal zone– using support vector machines as the fault classifier. These authors simulated three faults individually: a stuck damper, cooling coil fouling/block, and a decreasing supply fan speed, each at three levels of degradation. The authors use residual calculations between simulated normal operation and faulty operation. The results of model-based FDD proved highly accurate to detect and diagnose individual faults at varying degrees. Zhou et al. [40] developed a similar model-based FDD strategy for an entire HVAC system, including the cooling tower, chiller, pumps, and heat exchangers. The authors provide a method of using performance indices to evaluate the residual between normal and faulty operation. The performance indices evaluate the effect of faults within each sub-system. The author presents fault models of fan motor degradation, condenser and evaporator fouling, compressor motor degradation, partial clogs in pumps and pipes, and tube fouling, all of which are evaluated at two severity levels. The results indicate an ability to perform FDD accurately with increased severity levels. Finally, an ongoing study by the National Energy Renewable Laboratory investigates the application of fault modelling within HVAC simulations to develop FDD strategies [41]. Few papers include a discussion on boiler fault models, highlighting a gap within the research.

A handful of boiler-specific black-box and gray-box models were found in the literature. Romeo and Gareta [42] present the results of a feed forward neural network model capable of identifying the amount of fouling present in a biomass boiler. As well, the author describes a second method for fouling detection with a thermodynamic model. The authors then combine these two models and show a quicker and more robust ability to predict fouling within biomass boilers then each separately. Teruel et al. [43] present a methodology for detecting accumulation of ash on furnace through the application of feed forward neural networks and sensor data from heat flux meters. The author states that using several subsets of neural network allows the model to have increased physical meaning. However, the author notes the complexity of performing the proposed task. Alnami and Al-Kayiem [44] develop an automated FDD system using neural networks and collected data from a coal fired power plant. The authors were able to diagnosis super heater low temperature with a high degree of accuracy and at a speed faster than the plant monitoring system. Additional review of the literature indicates that modeling of boilers tends to focus on large-scale boilers commonly used in power plants; rather than those used in residential buildings.

The presented discussion of FDD is primarily focused on the theoretical applications of the domain knowledge necessary to develop a methodology capable of performing the task of FDD. Meanwhile, an FDD strategy is only as useful as its application towards actual building equipment. This application is further discussed in papers such as [45] who explore the challenges of integrating operation and maintenance facilities with Building Information Models. Additional research specific on integrating developed FDD methods for HVAC systems is explored in [46]. Therefore, a brief overview of the state of FDD has been presented.

## 2.4 Summary of Literature Review Findings

This review of the literature has shown that while boiler performance has been well-considered in the literature using physics-based approaches, there has been little application of FDD within the heating context. The papers that study FDD applications towards heating elements have been found to focus on industrial applications [31, 42, 43]. Therefore, research gaps are presented as such (1) there is a need for a validated white-box model capable of simulating boiler performance within the context of commercial/institutional buildings, (2) creation of a dataset for various operating conditions, encompassing fault-free and faulty performance, and (3) development and evaluation of a data-driven FDD tool capable of distinguishing between the various faults that may inhibit boiler performance. The importance of this research gap is found within legacy buildings, which lack the necessary data for the successful implementation of process history based FDD. In addition, the literature tends to focus on singular boilers, ultimately creating FDD tools for highly specific situations [12, 31, 43, 42].

This research paper addresses these gaps by presenting a data-driven boiler fault detection across a range of non-condensing boilers, with a focus on the machine learning algorithms most effective for individual and generalized fault detection and classification. The Simscape heating system model [22] described above was determined to be the most effective starting model based on those reviewed, in part because Simscape offers the ability to model both steady-state and dynamic performance, there have been various applications within research papers focused on modelling of heating [23, 24, 25, 26], as well as providing the fundamental components necessary to model this system. Simscape was used to emulate steady-state performance across nominal and faulty conditions, where the outputs were collected in dataset according to known BAS datapoints. The classification algorithms considered as potentially useful for this research include Random Forest [47], Decision Trees [48], and K-Nearest Neighbour [49] and Support Vector Machines [50]. The most promising trained models were then tested using a range of non-condensing boilers to test the generalizability of this approach.

## 3 Methodology

The methodology for developing an FDD tool for non-condensing boilers is presented. This research was undertaken in three steps: (1) development, calibration, and validation of a boiler emulator capable of simulating normal operation and operation under key fault conditions; (2)

simulation of fault condition dataset; and (3) creation of a testing and training dataset for a machine learning model to identify fault conditions based on standard building monitoring system data points. These three steps were repeated for a set of boilers representing a full range of sizes from a given manufacturer (the Viessmann Vitorond 200 series) as well as two boilers from a second manufacturer (Raypak) so that results across boiler sizes and manufacturers could be compared to evaluate the ability for the resultant classification algorithm to be generalized.

### 3.1    Emulator development, calibration and validation

Fundamentally, a steady-state boiler may be divided into two components as described in [51]: the combustion chamber and a heat exchanger. Simscape provides many of the components necessary for the development of this boiler, including a heat exchanger, water loop pump and a component that models combustion. These components were assembled to create the boiler emulator. The following discussion highlights fundamental first principle concepts necessary to ensure modelling accuracy.

The model assumes the combustion chamber is adiabatic, that complete combustion occurs in the normal operation case, that the input fuel (natural gas) can be assumed to be pure methane, and water vapor products do not condense. The products of reaction are assumed to include only carbon dioxide, water vapor, oxygen, and nitrogen; i.e. incomplete combustion is not permitted in this model in the normal case. The general combustion reaction equation with hydrocarbons is formulated in Eq. 1, where $x$ and $y$ signify the number of carbon and hydrogen molecules, respectively. Eq. 1 provides the fundamental chemical relationship between the inputs and outputs of the combustion chamber.

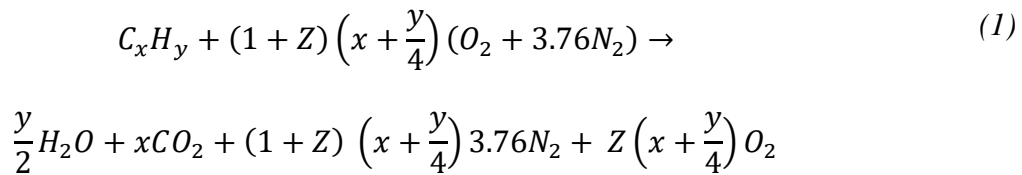

$$C_xH_y + (1 + Z)\left(x + \frac{y}{4}\right)(O_2 + 3.76N_2) \rightarrow \qquad (1)$$

$$\frac{y}{2}H_2O + xCO_2 + (1 + Z)\left(x + \frac{y}{4}\right)3.76N_2 + Z\left(x + \frac{y}{4}\right)O_2$$

The inputs of the combustion chamber include the mass flow rate of the fuel, the temperature of outdoor air, humidity ratio of outdoor air, and temperature of the fuel. The mass flow rate of the fuel was found using Eq. 2. The formula is used within *ANSI/AHRI Standard 1500: Performance Rating of Commercial Space Heating* [52]. Typically, manufacturers provide an input rating for their boilers, while the higher heating value (HHV) is a property attributed to the type of fuel. The flow rate of combustion gas establishes the magnitude of the heat generated through combustion.

$$\dot{m}_{gas} = \frac{Q_{in}}{HHV_{gas}} \qquad (1)$$

Properties of outdoor air were selected to reflect a relative humidity of 65% at various outdoor air temperatures. While the fuel temperature was left constant and unchanged from the default value of 303 K. Figure 2 depicts the boiler emulator developed in Simscape with the combustion chamber along with its inputs shown. Before calculating the output of the combustion chamber, parameters within the component must be defined to calculate outputs. Specific heat capacities, fuel heating values, and boiler geometries were assumed to be constant for each individual boiler, as summarized in Table 1. Note that while the inlet fluid temperature is assumed constant, this is due to the fact that this is true instantaneously and given the speed with which steady-state is achieved (typically less than 15 seconds, which is a typical-to-fast sampling rate for a BAS), this was determined to be an acceptable assumption for these models. A range of inlet temperatures was simulated to create the training dataset, thus capturing the longer-duration change in this parameter.

*Table 1. Summary of Constant Component Parameters Across All Boilers*

| Component | Parameter | Value and Source |
|---|---|---|
| Boiler | Hydrocarbon lower heating value | 50MJ/kg [53] |
| | Fuel specific heat at constant pressure | 2191 J/kg/K [53] |
| | Dry air specific heat at constant pressure | 1005 J/kg/K [54] |
| | Water vapor specific heat at constant pressure | 2900 J/kg/K [54] |
| Gas/Water Heat Exchanger | Flow arrangement | Shell and Tube |
| | Number of shell passes | 2 |
| | Wall thermal resistance | 0 K/W [55] |
| | Hydraulic diameter for pressure loss | 101.6 mm [56] |

| | Thermal Liquid - Initial temperature | 333 K [56] |
|---|---|---|
| Water Mass Flow Rate Source | Cross-sectional area at ports A and B | 0.1 m² [56] |
| PS Constant – Temperature Fuel | Constant (Temperature) | 303 K [56] |

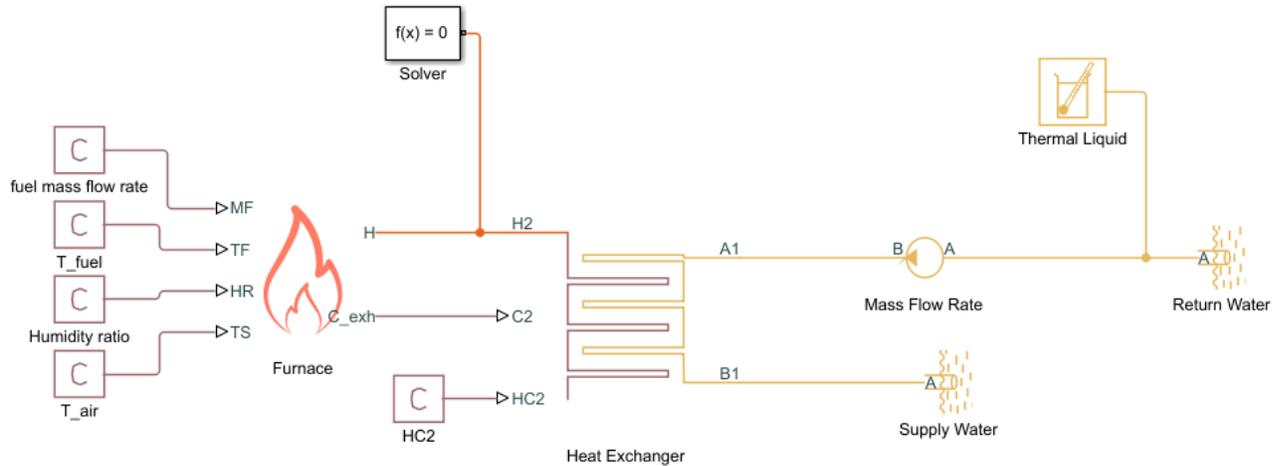

*Figure 2. Boiler emulator developed in Simscape.*

The only parameter that was necessary to calculate and validate was the Percent Excess Air or the (1+Z) term within Eq. 1. The value of excess air was validated according to the manufacturer specifications, often provided as either the amount of $CO_2$ or $O_2$ present in the flue gas. The combustion reaction results were replicated by calculating the mass percentage of the specified gas product. General expressions for these ratios are presented in Eq. 3 and 4.

$$\%CO_2 = \frac{xCO_2}{\left(\frac{y}{2}H_2O + xCO_2 + (1+Z)\,3.76N_2 + Z\left(x+\frac{y}{4}\right)O_2\right)} \qquad (3)$$

$$\%O_2 = \frac{Z\left(x+\frac{y}{4}\right)O_2}{\left(\frac{y}{2}H_2O + xCO_2 + (1+Z)\,3.76N_2 + Z\left(x+\frac{y}{4}\right)O_2\right)} \qquad (4)$$

The volume percentage of CO2 and O2 varies depending on the amount of excess air present during the reaction. The boilers were validated for the Vitorond boilers based on 10% CO2 present in the exhaust gas, as defined in the manufacturer specifications [56]. The outputs of the combustion chamber include combustion gas temperature, and heat capacity rate of combustion gas, and the heat generated through combustion. The temperature and heat capacity rate provide the necessary inputs to the combustion gas-water heat exchanger, while the heat generation is used to determine the efficiency of combustion.

A comparison between the theoretical flame temperature and the temperature output by the combustion chamber show a significant discrepancy. The values were 2330 K and 2755 K, respectively, roughly a difference of 18%. A non-condensing boiler exhausts gas containing water vapor, whereas the Simscape component chooses to omit the water vapor generated from combustion. Essentially, the model assumes that water vapor is condensing. A comparison between combustion gasses with and without water vapor is shown, along with the theoretical flame temperature in table 2.

*Table 2 Evaluation of Flame Temperature*

| Percentage Excess Air [%] | Theoretical Flame Temperature [K] | Combustion Gas Temperature (without vapor) [K] | Combustion Gas Temperature (with vapor) [K] |
|---|---|---|---|
| 0 | 2330 | 2755 | 2275 |
| 10 | 2192 | 2559 | 2145 |
| 20 | 2072 | 2392 | 2030 |
| 30 | 1967 | 2248 | 1929 |
| 40 | 1874 | 2122 | 1839 |
| 50 | 1791 | 2012 | 1758 |
| **Average Percent Difference** | | 15% | 1.51% |

The results clearly indicate that omitting water vapor is an incorrect assumption. Modelling the combustion gasses with water vapor is theoretically more accurate than not modelling with water vapor in the combustion gas. Therefore, the Simscape component was modified to include water vapor. The combustion chamber has been discussed in detail, now the discussion will continue on to the heat exchanger component.

The heat exchanger was assumed to be a two-pass shell-and-tube, with water in the tubes and combustion gas in the shell. The mathematical analysis used the effectiveness-NTU method [11]. Unlike the combustion chamber, the model parameters within the gas-water heat exchanger component are unique for each boiler model. The parameters that encompass these unique details of boilers include the thermal liquid volume, heat transfer surface area for the tube-side and shell-side, and heat transfer coefficients. Most of these values are found within manufacturer literature. For example, the literature for the Viessmann Vitorond 200 VD2 series provides thermal liquid volume and heat transfer surface area for each of their boiler models within the series, allowing the model to fit manufacturer specifications for 11 unique boiler models [56]. Liquid-side heat transfer coefficients are kept as the default value within the component and were assumed to be constant while the calculation of gas-side heat transfer coefficients is discussed further.

The importance of the noted parameters is due to the mathematical relationship when performing heat exchanger analysis. The equation for the overall heat transfer coefficient of a heat exchanger is presented in Eq. 5, which shows the fundamental relationship between the heat transfer coefficient and the geometry of the heat exchanger. Simscape implements these equations used for heat exchanger analysis, simplifying the modeling process. Regardless, inputs must be provided either from manufacturer data or through calculation.

$$UA = \frac{1}{h_i A_i} + \frac{R_{f,i}}{A_i} + \frac{ln\left(\frac{D_o}{D_i}\right)}{2\pi kL} + \frac{R_{f,o}}{A_o} + \frac{1}{h_o A_o} \qquad (5)$$

The inputs for the heat exchanger model include the thermal liquid side and the gas side areas and film and convection coefficients. The thermal liquid side inputs assume the liquid within the heat exchanger is water. The inputs include the mass flow rate of the water, return water temperature, and water pressure. Manufacturers provide these values since they provide the test conditions used for measuring the performance of boilers. The gas-side inputs include the outputs of the combustion chamber, the discussion of which has been presented. It should be noted that the overall areas are considered, rather than the detailed geometry of the tube and shell heat exchange surfaces. This is necessary because manufacturer data does not provide adequate detail for a more detailed geometric analysis.

A key parameter within the gas side of the heat exchanger includes a heat transfer coefficient of the combustion gases. Calculating the heat transfer coefficient was an iterative approach using manufacturer data of known operating conditions. The provided values of the temperature difference between supply and return water, the flow rate of the water, the heat output, the exhaust gas temperature, and combustion and thermal efficiencies are evaluated with each iteration of the heat transfer coefficient. A flow chart depicts the calculation of the heat transfer coefficient (Figure 3). The outputs reflect the operating conditions of the boiler model based on manufacturer data. The logic of the flowchart acts as a method for the calibration of the boiler. The model is calibrated to actual operating conditions and is able to accurately replicate its performance. The model was deemed valid once the simulated results were within an acceptable tolerance of the manufacturer's data. This method assumes that the heat transfer coefficient is constant throughout its operation. The results are further validated by simulating and evaluating performance of the model according to additional datapoints provided by the manufacturers. The results of the evaluation indicate that the assumption of a constant gas-side heat transfer coefficient is a valid one for each boiler over the operating range.

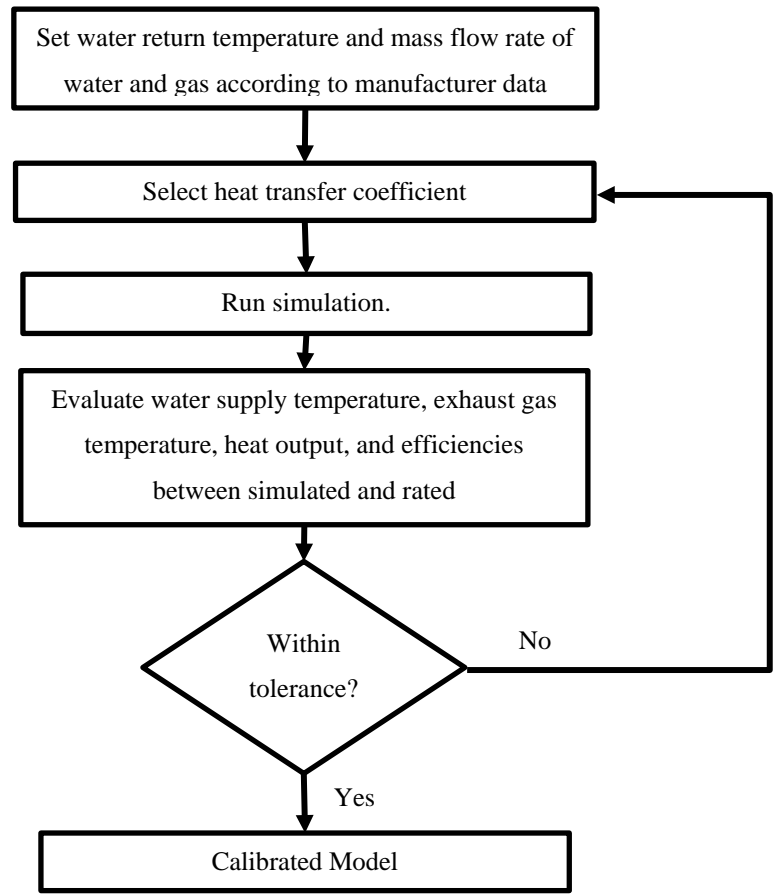

*Figure 3. The iterative approach used for calculating gas-side heat transfer coefficient*

The discussion has shown the methodology used to develop a model of a boiler using primarily manufacturer data and test conditions outlined by standards. The validity of the model was evaluated by comparing outputs published by the manufacturers to those simulated within Simscape. One final metric that evaluates the validity includes calculations of non-condensing combustion efficiency and thermal efficiency, as outlined in BTS-2000 [57]. First, non-condensing combustion efficiency calculates the efficiency of the boiler considering flue gas heat losses. Generally, the equation is formulated as:

$$Effy_{ss} = 100 - L_f \qquad (6)$$

$L_f$ represents the flue losses in the exhaust gas, which is a function of the amount of $CO_2$ present in the exhaust gas, the relative humidity of air supplied during combustion, the flue gas temperature, room temperature, and several provided constants [57]. To evaluate the combustion efficiency, it is necessary to calculate the temperature of the flue gas. This was performed by first obtaining the properties of water for the inlet and outlet and the properties of gas for the inlet, all of which are calculated by the Simscape component. It assumed no heat loss occurred; therefore, a heat balance between the three known states can be performed. This means the heat received by the water side is the same as the heat rejected by the gas side. Thus, the properties of the exhaust gas and the non-condensing combustion efficiency can be calculated. Finally, the thermal efficiency of the boiler is calculated considering the heat input and heat output. The heat input is found using Eq. 7 while the heat output is found through the simulation outputs. Thermal efficiency is thus calculated with Eq. 8.

$$Q_{in} = \dot{m}_{gas} \times HHV_{gas} \qquad (7)$$

$$\eta = \frac{Q_{out}}{Q_{in}} \qquad (8)$$

Therefore, the calibration of the boiler is achieved by evaluating the temperature of water and exhaust gas, products of exhaust gas, the flow rate of water, and thermal and non-condensing combustion efficiency. The boiler model was further validated by simulating additional flow rates and achieved temperature differences published by the manufacturer without changes to the calibration values.

### 3.2    Development of a classifier training dataset for FDD

In order to be practically useful, it is important that the most readily available parameters be used as input features for classifier training. To this end, the key control and sensor points from a BAS, namely entering and leaving water temperatures, pump VFD speed, outdoor air temperature, and room (fuel) ambient temperature. Initially the gas consumption rate was also considered, however this was difficult to measure in a series of case study buildings currently underway, and flue gas temperature measurement using an IoT-enabled thermocouple was found to be an effective alternative.

The purpose of this dataset is to train machine learning models to classify field data to detect or predict fault conditions. Because it is highly undesirable to run boilers in significant fault mode (extremely high excess air, or permitting boilers to operate after scaling and/or fouling are known to be present and significant), this is highly valuable to classify The dataset can be used to supplement field data and provide insight on the impact of graduated boiler faults on measurable points. While the authors agree that actual field data would be ideal, for boiler scaling and fouling measurements, such data would require the full disassembly of boilers at various points along the progression of this deterioration without remediating the problem. The lack of such data is thus understandable as this is infeasible and impractical. As forensic boiler fault data becomes available, however, this will serve critical to supplement this dataset and validate these simulations. In the interim, this simulated dataset allows the exploration of fault impact on boiler performance from first principles and uses the variables representing Building Automation System sensor points to develop fault detection and diagnosis algorithms.

A range of common boiler faults was incorporated through fault modelling: gas-side fouling, water-side scaling, and excess air. The performance of heat exchangers typically deteriorates due to the accumulation of deposits on the heat exchanger surface. Fouling increases the resistance to heat flow, therefore worsening the performance of the heat exchanger. This fault is represented as the fouling factor, $R_f$, the second and fourth term in Eq. 5. In practice, boilers operate with a certain amount of excess air for safety and maintenance reasons. Excess air is often introduced because it prevents incomplete combustion, the products of which include carbon monoxide, unburned fuels, soot, and surface fouling. However, as more excess air is introduced, the combustion efficiency decreases. An increased percentage of excess air causes a reaction to contain increased amounts of Nitrogen and Oxygen gas in the products, which absorb useful heat that is exhausted to the atmosphere, resulting in high losses. The modelling of excess air was performed by varying the inputs of equation 1, where term Z represents the percentage value of excess air. Fault modelling of excess air included simulating a range of excess air percentages.

Table 3 provides a summary of the ranges used for fault modelling. The rationale for the fault range considered is as-follows. The lowest case (1%) was used to determine the precision of the algorithm. The 5% steps up to 46% to determine how well the resultant FDD algorithms could detect progressive deterioration.

*Table 3. Summary of variables, faults, and emulator implementation*

| Fault (Label) | Component | Variable | Nominal Value | Tested Range |
|---|---|---|---|---|
| Excess air (X) | Boiler combustion chamber | Air flow rate | 0 | 1% - 46% |
| Gas-side fouling (F) | Gas-Water heat exchanger | Fouling factor (%) | 0 | F = 0.01 - 0.46 |
| Water-side Scaling (S) | Gas-Water heat exchanger | Scaling factor (%) | 0 | S = 0.01 - 0.46 |

To complement the fault data, the boiler model was simulated with a variety of normal operating conditions. The results of the simulations were used to generate a dataset intended to be used to analyse streamed data from a BAS as discussed previously. Iterations were performed, changing the gas fuel flow rate, water mass flow rate, outdoor air temperature, return water temperature, and simulating either normal performance or one of three faults and labelling the data with the fault class. This was performed for 14 non-condensing boilers resulting in datasets with thousands of simulations for each, which formed the testing and training datasets for FDD algorithm development.

### 3.3    Classification using machine learning

Several machine learning classification algorithms were tested on each of the generated datasets to create an FDD classifiers capable of analysing the input feature data and determining the nature of the fault. Consistent with best practices for machine learning research, a selection of common classification algorithms was tested for their ability to distinguish between normal operation and each fault, informed by the literature review. These were namely K-nearest neighbour (KNN), Decision Trees (DT), Random Forest (RF), and Support Vector Machines (SVM). The algorithms were programmed in Python using Scikit-learn [58], which was used for model training, testing, evaluation, and result visualization. The procedure of evaluating classification ability requires first splitting the data into 67% training/cross-validation set and withholding 33% for tuned algorithm testing. This split was consistent for all algorithms tested and a 5-fold cross-validation method was used.  Each machine learning algorithm was evaluated by performing a hyper-parameter search in conjunction with cross-validation, as summarized in Table 4. Grid Search was performed to evaluate an array of hyper-parameters which trained the algorithm with all possible combinations of parameters.

*Table 4 Hyperparameter Searches by Algorithm*

| Algorithm | Hyper-parameter tested | Range |
|---|---|---|
| K-nearest neighbour | K Number of Neighbours | {3,5,7} |
| Decision Trees | Maximum Depth | {5,10,15,20} |
| | Minimum Leaf Samples | {1,2,5,10} |
| | Minimum Split Samples | {2,5,10} |
| | Criterion | {Gini, Entropy} |
| | Splitter | {Random, Best} |
| Random Forest | Maximum Depth | {5,10,15,20} |
| | Minimum Leaf Samples | {1,2,5,10} |
| | Minimum Split Samples | {2,5,10} |
| | Number of Estimators | {50,100,150} |
| Support Vector Machines | Kernel | {Radial Basis Function} |
| | Gamma | {0.1,1 } |
| | C | {500, 1 000} |

The result of hyper-parameter tuning and cross-validation is a re-trained model with optimized classification results. The best hyper-parameters are applied to the testing data and classified. The result of the classified test data is evaluated through the scoring metrics: accuracy, precision, and recall.

The level of classifier precision was varied through this research. Initially, the FDD classifier was used to predict the exact fault category (i.e., fouling = 5%). Based on preliminary results, it was noted that there was substantial error between adjacent categories, particularly for excess air, and a revised approach was used where all excess air faults were labelled the same. This simplification of classes was further repeated until only the fault type but not the degree was classified.

### 3.4    Evaluation of Generalizability

The developed machine learning models were evaluated on their ability to perform classification on various iterations of datasets. The goal of this study is to assess the ability to generalize boiler performance by observing whether the machine learning models can retain a high classification accuracy despite inputting iterations of datasets. A model that remains high in accuracy would be considered generalized it were capable of not only distinguishing nominal and faulty performance of an individual boiler but also nominal and faulty boiler performance across multiple boilers. The evaluation of the models' ability to generalize is performed in several iterations, 1) training on mid-range boiler and testing on separate boilers of same manufacturer; 2) training on the entire dataset and testing on the entire dataset; 3) training on the entire dataset and testing on individual boilers; and 4) training on a set number of boiler models and testing on the remainder. Throughout each studies the hyperparameters were kept constant such that the most optimal hyperparameters are selected during the grid search. The training and testing were split 80% and 20% respectively for each study. The datasets used throughout these studies were the 22-class since their performance was found to provide the highest classification accuracy in the single-boiler models.

## 4    Emulator Calibration and Validation

A set of 14 non-condensing boilers was selected for this research, expanding upon an initial study considering a single boiler. These 14 included the full 11 models within the Viessmann Vitorond 200 boiler series, along with the Viessmann Vitogas 050 ECV-200, Raypak Raytherm 685-T, Raypak Raytherm H3-724. These models were created to support future case studies with local institutional and multi-unit residential building operators.

The emulator was simulated for each of the Viessman Vitorond 200 series models using manufacturer data to tailor the model parameters. Calibration was undertaken to minimize the error associated with five published outputs: thermal output, combustion efficiency, thermal efficiency, water supply temperature, and temperature of exhaust gas, all of which provided by the manufacturer according to AHRI standard tests [52, 57]. These outputs are gathered once the simulated boiler reaches steady-state operation. The accuracy of the simulated model is evaluated by comparing each of the outputs to the manufacturer specifications. A summary of the accuracy for each output is tabulated in Table 5 across all boiler sizes in this model. The efficiencies have an error below 3% and generally underestimate the efficiency of the boiler; this may be due to the "high efficiency three-pass design" of the heat exchanger; as noted in the emulator development section, only the total heat exchange area is published so it is possible that the true geometry provides better heat exchange than is assumed in the model. This would have a direct impact on the water supply temperature and the overall heat output, which are both also consistently underpredicted. In contrast, the exhaust gas temperature shows a high level of accuracy, with an average error of less than 0.5% and a deviation range of +/- 1°C. The 3% error was deemed acceptably low and deemed to be acceptable.

*Table 5. Calibration Error for Viessmann Vitorond 200 series*

|         | Combustion Efficiency | Thermal Efficiency | Exhaust Gas Temperature | Water Supply Temperature | Thermal Output |
|---------|-----------------------|--------------------|-------------------------|--------------------------|----------------|
| Average | 1.70%                 | 3.01%              | 0.53%                   | 0.496%                   | 3.59%          |
| Minimum | 1.36%                 | 2.84%              | 0.08%                   | 0.442%                   | 3.38%          |
| Maximum | 1.97%                 | 3.13%              | 1.26%                   | 0.514%                   | 3.86%          |

The remaining boiler types (Raypak Raytherm 685-T, Raypak Raytherm H3-724, and Viessmann Vitogas 050 ECV-200) did not have such data available for test conditions and did not show this deviation from manufacturer results, as illustrated in Figure 4. In these cases, the calibration accuracy of the simulations, proves to be highly accurate when compared to the data points provided by the manufacturer. The average percent validation errors of thermal output, thermal efficiency, and water supply temperature were 0.02%, 0.07%, and 0.02%, showing this calibration to be highly successful.

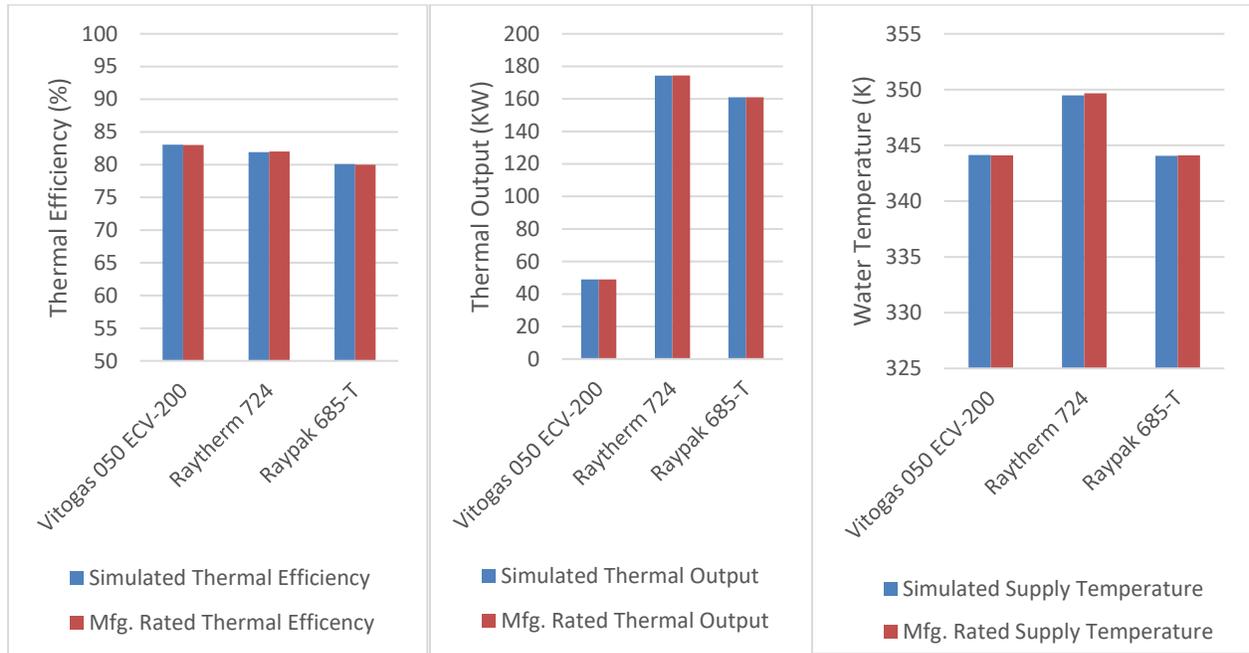

*Figure 4. Summary of Boiler Calibration for Three Models*

In addition to the capacity rating data, manufacturers provide additional datapoints of water temperatures at various water flow rates. Evaluating the performance of the simulation with these datapoints allowed the model to be validated against a broader set of empirical measurements. This is summarized in Table 5. Note that only the best and worst-performing Vitorond models are indicated in the table for brevity. The average error for the full set of boiler sizes was -0.16 K.

*Table 6. Results of Additional Datapoints Validation*

| Manufacturer/Model | Water Mass Flow Rate (kg/s) | Expected Delta Temperature (K) | Achieved Delta Temperature (K) | Error (K) |
|---|---|---|---|---|
| Vitorond 200 – 500 | 10.8 | 11 (20 F) | 10.8 | 0.21 |
| *(most error)* | 5.4 | 22 (40 F) | 21.5 | 0.46 |
| Vitorond 200 – 1080 | 23.2 | 11 (20 F) | 10.8 | 0.2 |
| *(least error)* | 11.6 | 22 (40 F) | 21.9 | 0.1 |
| Vitogas 050 | 1.05 | 11.1 (20 F) | 11.1 | 0.0 |
| | 0.69 | 16.6 (30 F) | 16.9 | 0.3 |
| Raypak 724 | 5.68 | 7.2 (13 F) | 7.3 | 0.1 |
| | 2.52 | 16.6 (30 F) | 16.5 | 0.1 |
| Raytherm 685-T | 6.31 | 5.5 (10 F) | 6.1 | 0.6 |
| | 3.47 | 11.1 (20 F) | 11.1 | 0.0 |
| | 2.33 | 16.6 (30 F) | 16.5 | 0.1 |

## 4.1    Energy Impact of Faults

Boiler faults can significantly affect heating outputs and efficiencies. Emulator results for the heat output impact of various faults were analysed to identify the scale of these impacts. Figure 5 provides an example of these impacts. This figure depicts two levels of each of three faults – air side fouling, water side scaling, and excess air – for the Vitorond 950 boiler. It is evident from this figure that while excess air has a relatively small impact on boiler performance (4% and 7% reductions in efficiency at 26% and 46% excess air), the decreased performance due to fouling and scaling are far more severe. As the boiler fouls, the heat output drops 63% at 26% fouled, and this increases to 78% at 46% fouled. From an energy use perspective, this is associate with an increase of 169% and 309% for gas consumption, with the same relative increase in $CO_2$ emissions to meet the same heating demand. Similarly, the increase from 26% to 46% scaling reduces heat output 76% and 86%, respectively, resulting in increases of 316% and 608% for gas consumption. When the other boiler outputs are compared, these trends are consistent.

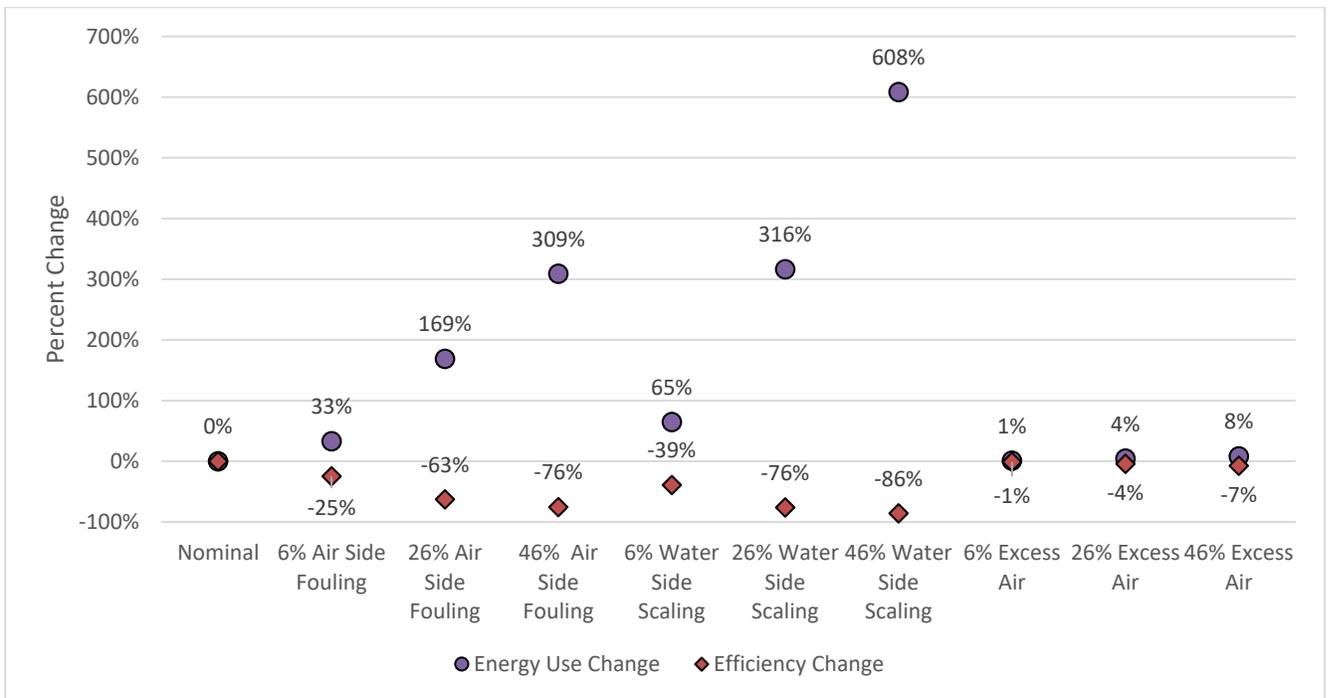

*Figure 5. Efficiency and Energy Consumption Changes due to faults for Vitorond 950 Boiler*

# 5   FDD Prediction Results

This section presents the results of applying machine learning classification as an FDD tool and evaluates the performance of each classifier across all boiler models. Each boiler model was simulated with constant input conditions without any faults present, thus generating the normal operation for each boiler. The following simulations observed the same input conditions; however, three unique fault models were incorporated. The fault models were simulated individually, which provides data for the progression of increasingly more detrimental operation. The outputs – fuel mass flow rate, the temperature of combustion air, water return and supply temperature, and water mass flow rate – were collected along with a label identifying the class of simulation (i.e., Normal, Fouling = 0.05, etc.). The datasets associated with the performance of each boiler is used as the basis for developing the FDD tool. The FDD tool was developed by using four machine learning classification algorithms: Decision Tree (DT), Random Forest (RF), K-Nearest Neighbour (KNN), and Support Vector Machines (SVM). The results of each classifier on each boiler are discussed, allowing each individual classifier to be evaluated as an FDD tool. Each machine learning classification algorithm can technically perform as an FDD since they each perform classification. However, accuracy must be preserved to ensure the FDD tool is practical. An accurate classifier is one that correctly labels data points for the entire dataset. Therefore, this discussion will evaluate the four machine learning algorithms with accuracy, referring to the ability of a classifier to label BAS data points correctly. Tables 7 and 8 summarize the prediction accuracy for each classifier and boiler.

*Table 7. Viessmann Vitorond FDD Accuracy Comparison*

| Classifier | Viessmann Vitorond Boiler Series Number | | | | | | | | | | |
|---|---|---|---|---|---|---|---|---|---|---|---|
| | **1080** | **950** | **860** | **780** | **700** | **630** | **560** | **500** | **440** | **380** | **320** |
| *KNN* | 62.0% | 71.6% | 72.4% | 73.4% | 74.3% | 75.0% | 75.6% | 75.3% | 74.4% | 77.5% | 77.2% |
| *DT* | **92.6%** | **93.8%** | **93.8%** | **94.8%** | **93.0%** | **92.2%** | **92.2%** | **91.8%** | **89.4%** | **92.6%** | **92.7%** |
| *RF* | 79.8% | 80.8% | 80.7% | 79.8% | 81.8% | 81.5% | 80.5% | 79.1% | 76.7% | 80.1% | 78.6% |
| *SVM* | 77.2% | 78.9% | 79.0% | 78.4% | 79.5% | 79.3% | 79.4% | 76.0% | 76.2% | 78.5% | 78.4% |

*Table 8. Raytherm and Viessmann FDD Accuracy Comparison*

| Classifier | Boiler Model | | |
|---|---|---|---|
| | **Raypak 685-T** | **Raypak 724** | **Vitogas 050 – ECV 200** |
| *KNN* | 81.0% | 87.9% | 92.4% |
| *DT* | **93.4%** | **95.4%** | **97.2%** |
| *RF* | 81.9% | 86.3% | 92.8% |
| *SVM* | 77.9% | 82.3% | 69.6% |

The classification accuracy of each model is investigated further through observation of the confusion matrices. The results of classification accuracy remain consistent between each of the 14 boiler models; therefore, the following discussion will focus only on a few boilers to remain concise. Firstly, the most prominent issue observed is the confounding of the excess air categories across each boiler within the DT, RF and SVM classifiers. As seen in Figure 6, adjacent categories of excess air are often mislabeled. This is most likely a result of the boiler performance varying minimally with small changes in excess air percentage. Furthermore, small percentages of water scaling, air fouling, and excess air are often misclassified by RF, DT, and SVM classifiers, again because their effect on boiler performance at these levels is limited. The results show that normal performance is classified perfectly throughout the DT, RF, and SVM classifiers. Overall, faulty performance is distinctly separated from normal performance. Upon evaluation of the results of each classifier, modifications to the dataset were performed to improve prediction accuracy.

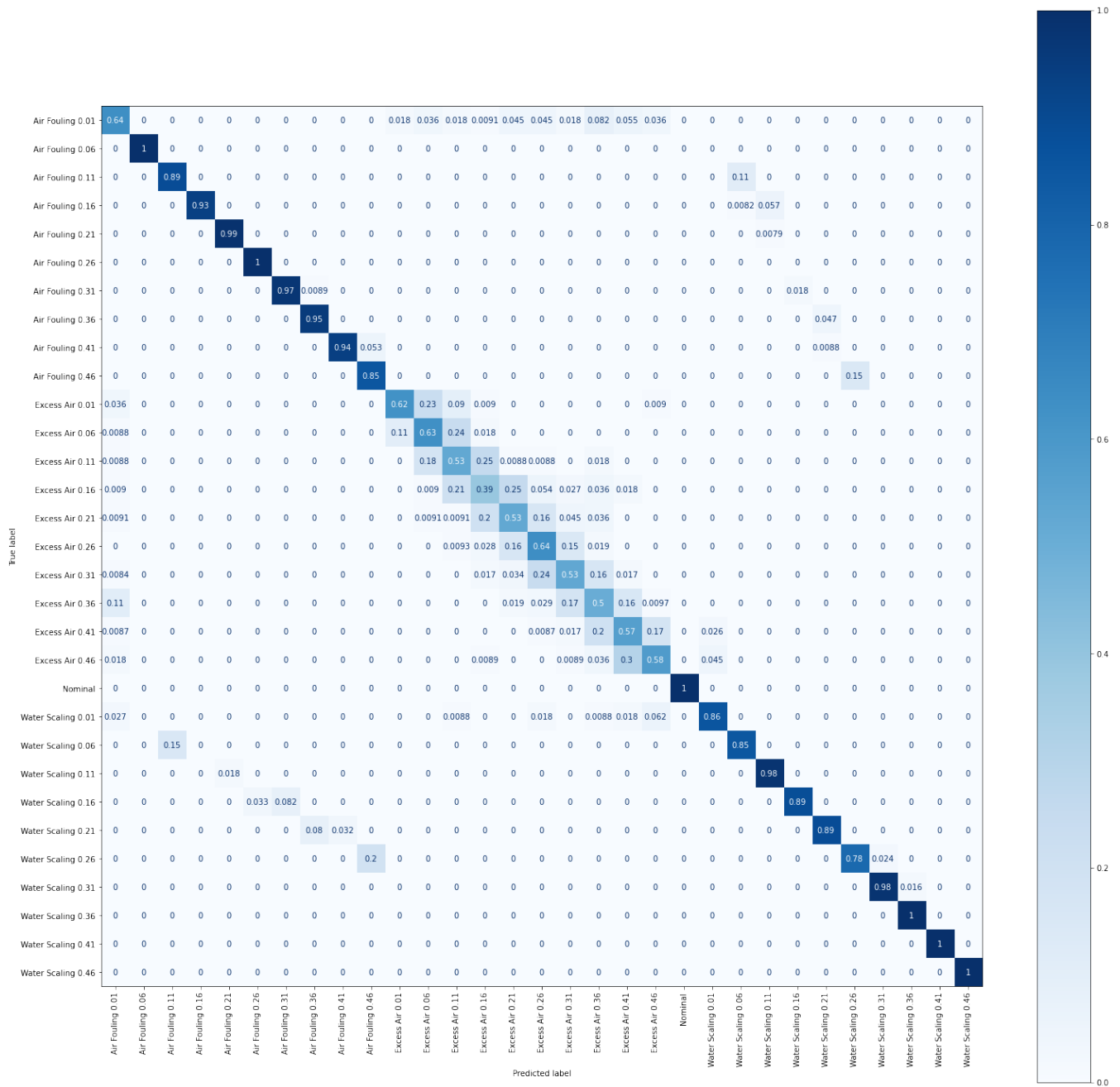

*Figure 6. Vitorond 950 - Random Forest {'bootstrap': True, 'max_depth': 20, 'min_samples_leaf': 1, 'min_samples_split': 5, 'n_estimators': 150}*

Modifying the datasets to improve accuracy was performed first by categorizing excess air as a single class. This was justified by understanding that excess air faults result from stuck outdoor air dampers and this is not a problem that is known to progress over time. In this classification scheme, all cases of excessive air percentages were labeled as "Excess Air". The results were a significant improvement in accuracy. Figure 7 depicts the results of random forest for the Vitorond 950 with a modified dataset, the accuracy of which is 95.1%, an increase of 14.6%.

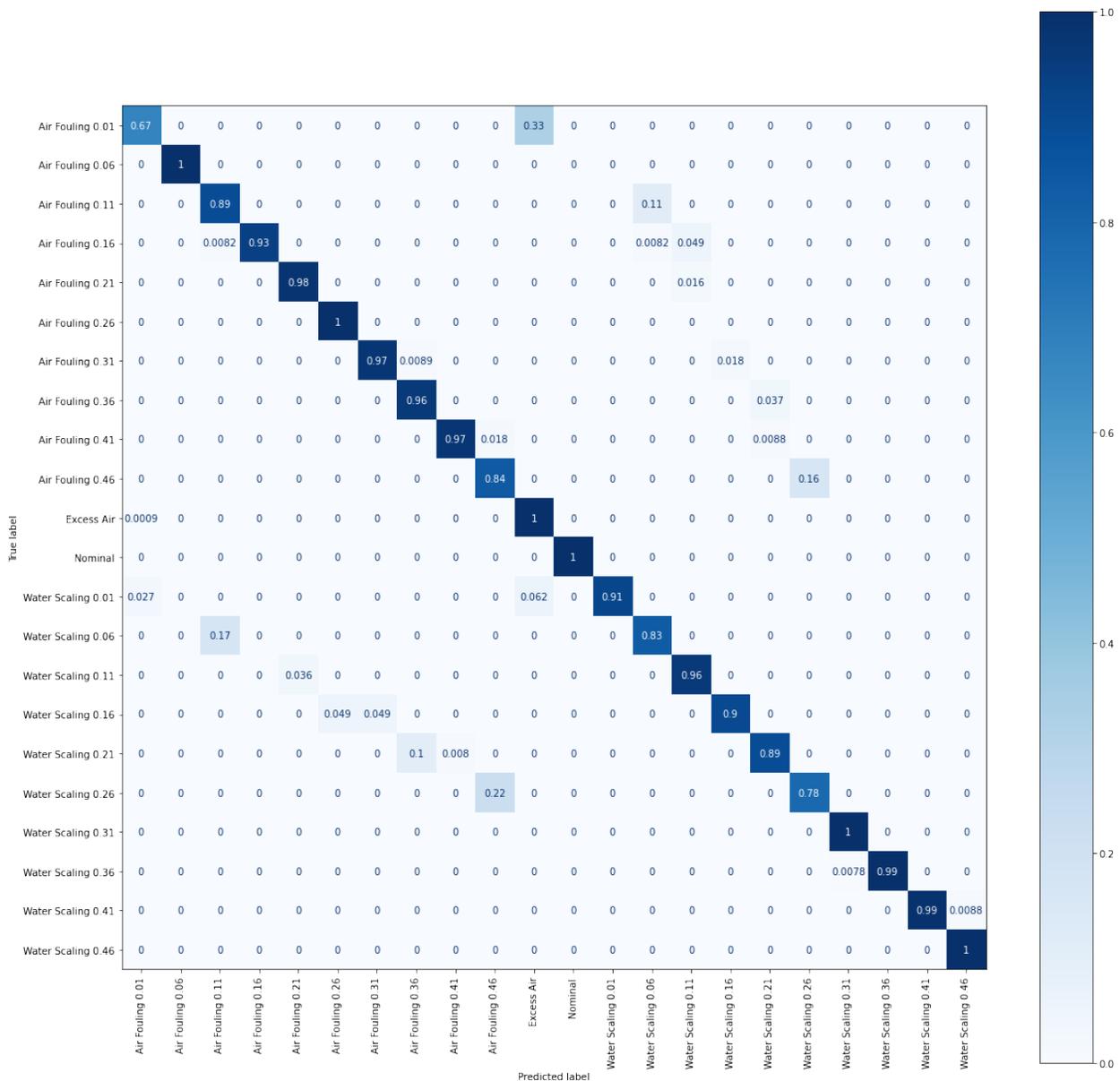

*Figure 7. Vitorond 950 – Random Forest with Single Excess Air Class {'bootstrap': True, 'max_depth': 20, 'min_samples_leaf': 1, 'min_samples_split': 2, 'n_estimators': 150}*

Following the discussed modification, the effect of broadening each class label was observed. Instead of 31 detailed class labels, 4 broad categories are provided, where classification is performed with each fault category. The goal is to have misclassifications between adjacent faults of the same type ignored, ideally further improving accuracy. The observed results for the categorical classification with random forest classifier on Vitorond 950 are presented in Figure 8, the accuracy of which is 91.7%, an increase of 10.9% from the primary classification model.

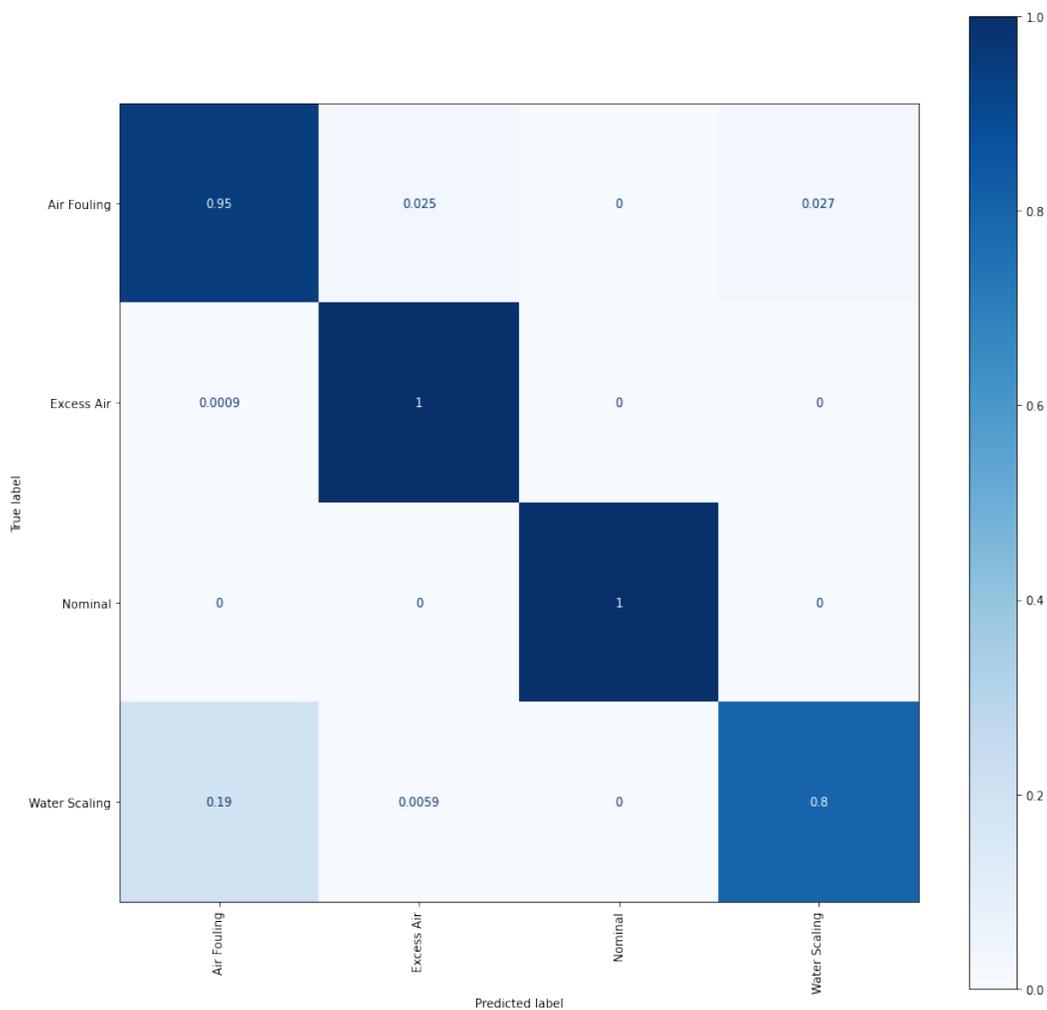

*Figure 8. Vitorond 950 – Random Forest with 4-class classification {'bootstrap': True, 'Maximum Depth': 20, 'Minimum Leaf Samples': 1, 'Minimum Split Samples': 2, 'Number of Estimators': 150}*

The categorial classification scheme proved unnecessary as the change in accuracy was minimal in comparison to the first modification. It was assumed preventing inter-class misclassification would improve accuracy; however, their impact on overall accuracy is limited. As well, the accuracy of the SVM classifier was impacted severally by the categorical classification. The full condition classification allows for increased granularity in prediction. This is important because it permits a more precise diagnosis of the specific fault occurring within the boiler. Therefore, the overall accuracy of a categorical classification proves to be poorly suited for FDD.

A summary of the results for several boiler models after each dataset modification are presented in the following tables. Due to the consistently poor performance of KNN, its are omitted.

*Table 9. Single Excess Air Categorization (22 Class) Accuracy Summary*

| Classifier | Viessmann Vitorond | | | | | Raypak | | Viessmann Vitogas 050 |
|---|---|---|---|---|---|---|---|---|
| | 1080 | 950 | 560 | 440 | 380 | 685-T | H3-724 | ECV 200 |
| DT | 96.8% | 97.2% | 96.6% | 97.6% | 96.8% | 98.2% | 96.8% | 96.2% |
| RF | 93.5% | 95.1% | 94.3% | 94.1% | 94.0% | 94.3% | 93.9% | 93.2% |
| SVM | 92.3% | 93.9% | 93.9% | 93.6% | 94.0% | 86.2% | 93.8% | 91.7% |

*Table 10. Categorical Classification (4 Class) Accuracy Summary*

| Classifier | Viessmann Vitorond | | | | | Raypak | | Viessmann Vitogas 050 ECV 200 |
|---|---|---|---|---|---|---|---|---|
| | 1080 | 950 | 560 | 440 | 380 | 685-T | H3-724 | |
| DT | 96.4% | 98.0% | 97.7% | 96.5% | 94.2% | 96.3% | 96.9% | 97.3% |
| RF | 91.7% | 91.7% | 90.8% | 91.8% | 91.7% | 90.4% | 91.2% | 93.2% |
| SVM | 77.1% | 77.0% | 77.3% | 76.9% | 77.1% | 75.8% | 76.7% | 72.0% |

Evaluating the hyperparameters after modifying the dataset shows similar parameter selection to those of the unmodified dataset. DT exhibited a similar issue of varied 'criterion' selection for each boiler model. RF varied with the 'number of estimators' and 'minimum samples split,' while 'minimum leaf samples' remained constant. SVM exhibited constant hyperparameter selection with a 'C': 1000 and 'Gamma': 0.1, across all boiler models. The best performing FDD scheme, 22-class modification, provided an overall model accuracy of 94%. DT generally performed with the highest overall accuracy of each classifier, an average of 97%. However, the high accuracy of DT may be a result of overfitting; the average accuracy of RF – 93% - remains high while avoiding this overfitting risk.

## 5.1    Generalization

One of the benefits of observing multiple boiler models and their performance with the various machine learning classifiers was the potential to evaluate their generalization potential. Two aspects of generalization were considered – first, the form and hyperparameters of the individual models, and second, the ability to iteratively manipulate the dataset to observe performance of single boilers. In the first case, a degree of generalization was observed using the 22-class boiler model. The hyperparameters remained similar throughout all the boilers and each classifier. The SVM classifier and its hyperparameters were consistent (Gamma = 0.1 and C = 1000). The hyperparameters of DT remained the same (Criterion = entropy, Maximum Depth = 20, Minimum Samples Leaf = 1, Minimum Samples Split = 2, Splitter = Random). Finally, the hyperparameters of RF remained the same (Bootstrap = True, Maximum Depth = 20, Minimum Samples Leaf = 1, Minimum Samples Split = 2, Number of Estimators = 150). This suggests that this methodology can be readily applied to other non-condensing boilers with the expectation of high prediction accuracy.

The second study observed four iterations of manipulated boiler datasets. This study sought to evaluate the ability to generalize individual boiler performance accurately. A generalized boiler FDD approach would allow this methodology to be extended towards new unlabeled boiler models, without the need for the development of a training dataset using the emulator. The four iterations include: 1) training on one boiler and testing on similar sized boilers; 2) collecting all the boiler datasets, training on the collected dataset and testing on individual boiler datasets; 3) collecting all the boiler datasets, training on 80% of the collected dataset, and testing on the remaining 20% of the collected dataset; and 4) each classifier was trained on all boilers except the Vitorond 950, 630, 500, 380, Raypak 685 and Vitogas 050, while the testing was performed on the withheld boiler datasets. Table 11 summarizes the results of the classification models. The classifier accuracies for iterations 1 and 4 are summarized for the algorithms with the highest achieved accuracy (SVM, DT and RF, respectively) while iteration 2 shows the average accuracy across all boiler models tested.

*Table 11. Boiler Model Generalization Results*

| Iteration Number | Training Input | Testing Input | Classifier Test Accuracy |
|---|---|---|---|
| 1 | Vitorond 560 | Vitorond 1080 | 48.8% |
| | | Vitorond 630 | 67.8% |
| | | Vitorond 500 | 75.0% |
| | | Vitorond 440 | 49.8% |
| | | Vitorond 320 | 38.8% |
| 2 | Collected boiler dataset | Individual boilers tested one at a time | **89.6% DT**[*] |
| | | | 88.6% RF[*] |
| | | | 58.3% SVM[*] |
| 3 | Collected boiler dataset | Collected boiler dataset | 91.4% DT |
| | | | 90.2% RF |
| | | | 61.6% SVM |
| 4 | Vitorond 1080, 950, 860, 780, 700, 560, 440, 320 Raypak 724 Vitogas 050 | Vitorond 630 Vitorond 500 Vitorond 380 Raypak 685 | 50.2% SVM 48.4% SVM 46.2% SVM 35.6% SVM |

[*]indicates average accuracy across all boiler models tested

The results of the various iterations generally show poor ability to generalize. The first study indicates that training on a single boiler was not applicable to predict fault in boilers of similar sizes from the same manufacturers. It was observed that the results of classification are accurate for classification of excess air, the test results distinguished nominal and excess air reasonably accurately with minor confounding between the two classes. However, many of the inaccuracies are a result of misclassification of fouling and scaling with no noticeable trends in the confusion matrices for the Vitorond 1080 and 320. In contrast, the relative best performers, Vitorond 500 and Vitorond 630, with both cross-fault error (i.e. low rates of fouling confused with low rates of scaling) and intra-fault error (confounding between different levels of the same fault), however normal and excess air were well-distinguished, as were the higher degrees of scaling. This improved performance may be a result of an overlap in performance characteristics between the training and testing input, namely the similarity boiler geometry, heating output, and mass flow rates for gas and water. However, the accuracy of the classifier shows that the overlap is insufficient to perform accurate FDD. The second iteration showed promising results, with accuracy that is generally high across the range of boilers tested. The disadvantage of this approach is the requirement for a collected dataset. If the collected dataset is missing a specific boiler, the results may worsen. This hypothesis is tested in iteration 4 where datasets are purposefully removed from the collected training input. Iteration 3 shows similar results to the second iteration. This may be a result of training the classifier with the same dataset and hyperparameters. Despite promising results, the same disadvantage as iteration 2 exists. Finally, iteration 4 evaluated the ability to classify with withheld boiler datasets. The results of this iteration were poor with consistently poor results. Again, the mid-range models (Vitorond 500 and 630) had better performance relative to the others but still fails the generalization test. These tests demonstrate the limited generalization of boiler models. While limiting the applicability of any individual boiler model, this shows the significant value of the emulator to develop the necessary datasets to expand this FDD approach to future boilers.

## 6 Discussion and Conclusions

The application of machine learning classifiers to boiler FDD showed several consistent trends. First, a comparison of prediction accuracy between the algorithms tested showed that DT consistently outperformed all other classifiers across the boiler models for the specific fault classification (31, 22, and 4 classes), and thus provides the most robust approach for the fault detection for this heating system. However, it is noted that DT exhibits signs of overfitting due to the decrease in accuracy of RF. Despite this change, RF performs with a high degree of accuracy with 22 and 4 class datasets. In addition, SVM shares a similar degree of accuracy for only the 22-class dataset. The 4-class dataset resulted in poor accuracy when applied with SVM. KNN had by far the poorest performance, showing it to be inappropriate for this type of classification. Further, most of the classifiers had consistent hyperparameters; showing a trend of generalizability for individual boiler models.

The three approaches to fault labeling also showed significant differences and provided insight on future work of this kind. The initial (31-class) dataset considered not only the fault type but the degree (in 5% increments) and only moderate accuracy (85%) was achieved. A closer look at the results indicated that the most significant source of error was the inability for the classifiers to distinguish between adjacent classes, most significantly for excess air. This resulted in high confounding of adjacent classes, for example 6% vs 11% but very little confounding between more distinct classes. Similar confounding occurred for moderate values of water-side scaling and air-side fouling. Despite the misclassification, each classifier had exceptionally high accuracy with mid-to-high levels of fouling/scaling. This ability to distinguish between degrees of fouling and scaling is particularly valuable as this will allow building operators to track the progression of these issues and quickly identify if a problem is progressing slowly or rapidly and plan equipment maintenance in response. The ability to distinguish between various degrees of excess air is less critical, as this issue is typically not one that progresses over time. In addition, DT, RF, and SVM were able to classify normal performance with 100% accuracy, showing that normal and faulty performance is distinctly separated.

To improve algorithm accuracy while recognizing the limited value of distinguishing between amounts of excess air, the data was relabeled to 22 classes: all distinction between fouling and scaling were retained but excess air faults were considered as a single class. This resulted in a significant increase in overall classification accuracy: SVM increased to an average of 93% prediction accuracy from 78% and DT and RF increased from 93% to 97% and 80% to 93%, respectively. Even KNN, which had an accuracy 74% initially increased to 85%. Once again, the hyperparameters across all classifiers were constant across all boiler models.

The final modification sought to eliminate the misclassification within boiler classes by assembling the range of faults as a singular class, resulting in a simple 4-class dataset. This resulted in reasonably consistent results in RF and DT prediction accuracy compared with the 22-class labels, but the accuracy of SVM suffered significantly. The KNN classifier had increased slightly in accuracy, outperforming SVM. This approach was deemed to be the least useful of all, due to the inability to track fault progression over time. Overall, the 22-class approach was found to be the most effective to support an intelligent boiler monitoring system for integration into the building automation system. This will provide operators with a previously unavailable degree of insight into boiler fault progression, allowing for improved maintenance schedules and permitting the optimization of operational costs.

This research had also explored the possibility to generalize boiler performance. The goal of generalizing boiler performance includes achieving high levels of classification accuracy despite missing or new data. Four studies were performed as metrics for generalizability of the boiler datasets. However, the results were generally poor across all the studies that relied on the machine learning algorithm to infer boiler performance. The classifier was unable to provide results with high classification accuracy because the trained model was incapable of generalizing the performance across nominal or faulty performance. Therefore, this highlights the importance of the boiler emulator within the proposed FDD strategy to provide accurate classification results.

The development of this FDD tool presents promising results, although there are several limitations of this research as presented. Firstly, the classification is only performed for individual faults, not combined/hybrid faults. Second, this research presents only simulated fault results and manufacturer-provided nominal data; the inclusion of field-collected data would both serve to validate the fault results (where such field data is available) and extend the database with more complex real-world conditions. Addressing the first limitation, multiple concurrent faults will be simulated to permit more complex investigations to be undertaken. To address the second limitation, the authors are obtaining real data from in-situ boilers, which will be used to both enhance the dataset as well as further refine and validate the fault detection models, once such faults are detected and manually diagnosed. There are significant limitations to the fault data that can be collected, however. First, low-level fouling and scaling can be difficult to identify from decreased boiler performance and therefore those data points would likely be obtained by chance inspections. Second, the degree of fouling and scaling would be difficult to quantify manually and may require much longer out-of-service times than is acceptable to a facility engineer. Third, fouling and scaling cannot be permitted to progress to the levels simulated without risking catastrophic damage to the boiler and potentially other heating system elements, which would require testing to failure in a lab, rather than field, setting. For these reasons, there is value in using simulated data for FDD training to provide the full range of points not possible to obtain or simply not available from such field studies. Finally, while typical boiler control and monitoring points have been selected for use, there will be installations where some are not readily available. Feature analysis would provide insight on the most valuable points for optimal FDD performance.

Beyond the scope of this paper, additional research is underway to investigate the impact of signal noise on prediction accuracy, identify signal processing techniques to increase the robustness of the model for real-world applications, and develop pre-processing algorithms for the time-series data collected from the field sites to use with the machine learning classifiers. Examples of applications of filtering and signal condition techniques used to condition noisy sensory data and remove unwanted components are discussed in [59]. These techniques have been successfully implemented in hardware and software. For example, a low-pass filter is usually adopted directly after a sensor to attenuate any unwanted high-frequency components injected from the surrounding environment. A practical application of digital signal processing techniques used to improve deteriorated signals caused by outliers is depicted in Figure 9. The figure shows a real outdoor air temperature signal that suffers from sharp spikes and the conditioned signal after applying a median filter.

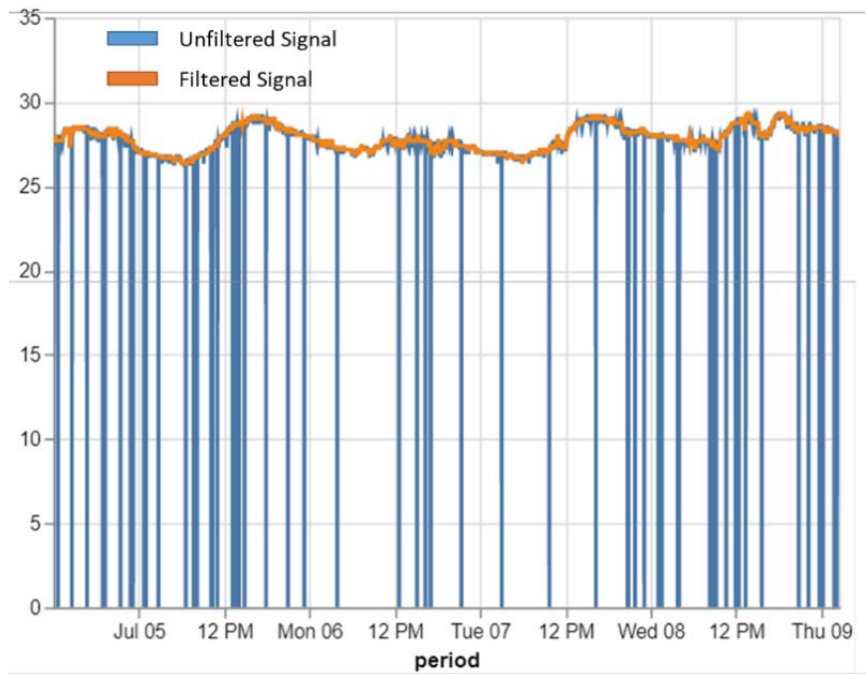

*Figure 9 Outdoor Air Temperature Signal Unfiltered and Filtered*

Regarding faulty and/or missing readings, Kalman and Bayesian filters [60] can be applied to improve the robustness of the implemented algorithm by estimating the required signals either from other sensory data such as the voltage and current of the monitored equipment or from previous records. Further comprehensive treatment of the practical implementation issues will be the subject of a future manuscript.

The presented work provides a novel approach for the design and evaluation of FDD algorithms used for non-condensing boilers, commonly found within legacy buildings. MATLAB/Simscape emulators were developed and validated for 14 different boiler models and used to generate fault data for training FDD classifiers. The most successful classifier used consistent hyperparameters and successfully detected fault conditions using only data accessible to standard legacy building automation systems, demonstrating significant potential for scaled deployment. This work contributes to the theoretical development of FDD approaches by utilizing a physics-based boiler model for data generation of normal and faulty performance. The developed FDD methodology is a simulated data driven FDD that captures faulty performance through fundamental physical laws.

This boiler emulator provides increased control on many of the parameters, inputs, and outputs that dictate the boiler performance and data collected, a control that is lacking within actual HVAC equipment. This allowed the FDD approach to be evaluated with varying outputs, providing the optimal theoretical strategy for non-condensing boiler FDD development. The results, an average classification accuracy across all boiler models of 94.0%, indicates the successful theoretical FDD framework for boiler FDD. In addition, despite the lack of sensor and IoT data, this research still proved valuable for maintenance purposes. The value includes generating data that represent faults that cannot be determined without removing boilers from service and thus such datasets do not exist. As well, this research acted as a case study by modelling 14 non-condensing boilers across a range of sizes and manufacturers. The calibration and validation methodology provided sufficiently accurate results that conformed to steady state performance published by manufacturers and boiler standards. The robust calibration methodology provided a basis for dataset generation that was further leveraged for FDD development. Finally, this research provides operators with the ability to leverage BAS data to detect and track boiler degradation or poor operating characteristics. These boiler operators will be equipped to rapidly correct issues, thus maintain a high level of boiler efficiency and maximizing service life.

## Acknowledgments

This research was funded by the Natural Science and Engineering Research Council (NSERC) Discovery Grant [RGPIN-2018-04105 and DGECR-2018-00395] and the Mitacs Globalink Internship Program.